\documentclass[pra,showpacs,twocolumn,superscriptaddress,amsmath,amssymb]{revtex4-1}

\usepackage{graphicx} 
\usepackage{dcolumn}

\begin{document}

\title{The Geometric Phase Appears in the Ultracold Hydrogen Exchange Reaction}
\author{B. K. Kendrick}
\affiliation{Theoretical Division (T-1, MS B221), Los Alamos National Laboratory, Los Alamos,
New Mexico 87545, USA}
\email{Correspondence should be addressed to BKK (bkendric@lanl.gov).}
\author{Jisha Hazra}
\author{N. Balakrishnan} 
\affiliation{Department of Chemistry, University of Nevada, Las Vegas, Nevada 89154, USA}
\vskip 10pt
\begin{abstract}
Quantum reactive scattering calculations for the hydrogen exchange reaction
H + H$_2$($v=4$, $j=0$) $\to$ H + H$_2$($v'$, $j'$) and its isotopic analogues 
are reported for ultracold collision energies.
Due to the unique properties associated with ultracold collisions,
it is shown that the geometric phase effectively controls the reactivity.
The rotationally resolved rate coefficients computed with and without the geometric phase are shown 
to differ by up to four orders of magnitude. The effect is also significant in the vibrationally
resolved and total rate coefficients.  
The dynamical origin of the effect is discussed and the large geometric phase effect reported here might be exploited to control
the reactivity through the application of external fields or by the selection of a particular nuclear spin state.
\end{abstract}

\maketitle


In the Born-Oppenheimer description of molecules the 
electronic Schr\"odinger equation is solved to obtain an effective 
potential energy surface (PES) which is then used in the solution of the
nuclear motion Schr\"odinger equation.
The electronic PES often becomes degenerate with an excited electronic state 
resulting in a conical intersection (CI). 
As noted long ago by Longuet-Higgins\cite{LH58} and Herzberg and Longuet-Higgins,\cite{herzberg63} the
electronic wave functions associated with a CI change 
sign for any nuclear motion pathway which encircles the 
CI (i.e., they are double-valued).  
The electronic sign change implies that a corresponding sign change
must also occur on the nuclear motion wave function.
Mead and Truhlar\cite{meadtruhlar79} showed that 
this can be accomplished by including an effective vector potential in the
nuclear motion Hamiltonian.  
Mead\cite{meadBS} originally referred to this effect
as the ``Molecular Aharonov-Bohm'' effect
but it is now commonly referred to as the ``geometric phase'' or ``Berry's phase'' effect.\cite{berry84,arno}

The most studied of all chemical reactions is the hydrogen exchange
reaction H + H$_2$ $\to$ H + H$_2$ and its isotopic analogues 
H + HD $\leftrightarrow$ D + H$_2$
and D + HD $\leftrightarrow$ H + D$_2$.
The H$_3$ system exhibits a CI between the ground
and first excited electronic states for equilateral triangle ($i.e., D_{3h}$) 
geometries.
As first predicted by Mead, the GP alters the relative
sign between the reactive and non-reactive scattering amplitudes for
the H + H$_2$ reaction which significantly alters the angular dependence 
of the differential cross sections (DCSs).\cite{meadScatter,kuppermannJ01990}
Unfortunately, state-resolved experiments for H + H$_2$ 
are very difficult in practice and Mead's predictions have not yet been verified. 
Though the isotopic variants are more accessible experimentally,
theoretical calculations showed negligible GP effects 
for a wide range of collision energies.\cite{kendrickHD2GPJ5,kendrickHD2GPJ,kendrickDH2GPJ,kendrickFeature2003,carlosAlthorpe2005,Althorpescience2005,Althorpe2006,Althorpe2008,Boukline2014}
Some relatively small rapidly varying oscillations in the DCS due to the GP have been seen in the 
theoretical DCSs at energies below that of the CI.\cite{carlosAlthorpe2005,bouaklineGP_2008,BoulkinH3}
At energies above the CI, large GP effects on the DCS's were observed which give rise to broader bi-modal features.\cite{BoulkinH3,bouaklineGP_2008,bouaklineQCT_2010}
However, GP effects remained elusive in the integral cross sections or reaction rates at any energy.
A recent experimental attempt to measure the GP oscillations in the DCSs for
the H + HD $\to$ H + HD reaction at energies below the CI was unsuccessful.\cite{zare2013}

Until recently,\cite{GPHO2cold2015}
all previous theoretical predictions of GP effects on chemical reactivity and 
experimental attempts at its detection have been done at thermal energies.  
Recent experimental progress in the cooling and trapping of molecules presents a novel energy regime at 
sub-Kelvin temperatures to explore GP effects in chemical reactions.\cite{carr09,ospelkaus2010}
In the zero-temperature limit where only $s$-wave contributes, the reaction rates obey the well known 
Bethe-Wigner threshold laws and approach finite measurable values.\cite{Bethe35,Wigner48}
In this Letter, it is shown that due to the unique properties associated
with ultracold collisions, namely:  (1) isotropic scattering,
and (2) quantized scattering phase shifts,
the maximal possible interference between the different scattering pathways 
around a CI becomes possible.
The maximal interference effects are shown to occur in the fundamental
hydrogen exchange reaction which results in very large GP effects
effectively turning on or off the reactivity.
The quantized scattering phase shifts are a 
general property of ultracold collisions and can occur
for interaction potentials which support bound states\cite{GPHO2cold2015}
as well as those which do not (as demonstrated in this Letter for the H$_3$ system).
In the latter case, suitable vibrational excitation of the reactant diatomic molecule
is required which results in an effective reaction pathway (along the vibrational adiabat)
that is barrierless\cite{cote2011} and exhibits a potential well.\cite{miller1980,truhlar1996,jankunas_PNAS2014}

\begin{figure}[h]
\includegraphics[scale=0.2]{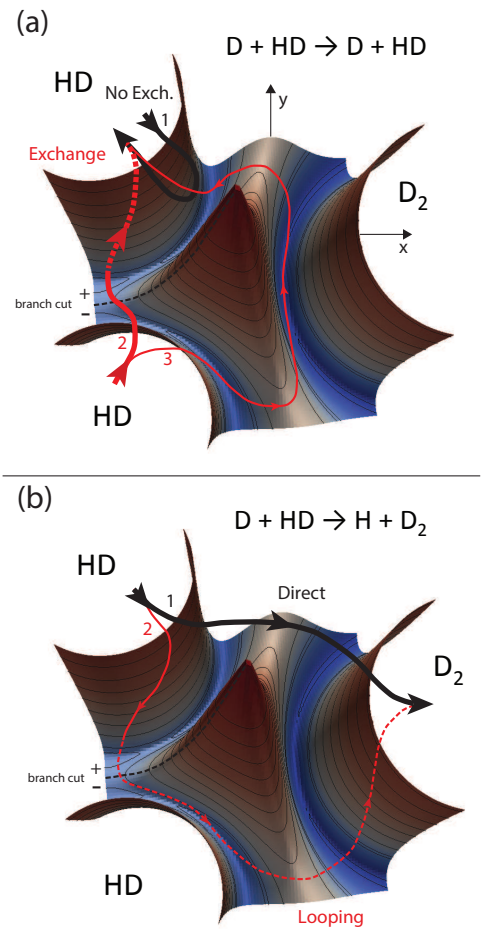}
\caption{
\begin{small} {(color online) A 2D slice of the 3D Born-Oppenheimer PES for the HD$_2$ system
is plotted at a fixed hyperradius $\rho=3.75\,{\rm a_o}$.  The different scattering
pathways around the CI are indicated, panel (a) for D + HD $\to$ D + HD, and
panel (b) for D + HD $\to$ H + D$_2$. For clarity, pathways for only one of the
symmetric HD channels is depicted (see text for discussion).}
\end{small}
}
\label{fig1PES}
\end{figure}

Figure \ref{fig1PES} plots a 2D slice of the 3D ground state H$_3$ electronic PES\cite{BKMP2} 
and reaction pathways for the D + HD $\to$ D + HD (panel a) and 
D + HD $\to$ H + D$_2$ (panel b) reactions.
Hyperspherical coordinates are used which have the advantage of showing
all arrangement channels simultaneously as well as the prominent CI
located near the center of the plot.\cite{Pack87}
Figure \ref{fig1PES} corresponds to a stereographic projection of the 
upper half of the hypersphere with a fixed hyperradius of $\rho=3.75\,{\rm a_o}$.
The zero of energy is the bottom of the asymptotic H + D$_2$ potential well.
The contour lines are separated by $2,900\,{\rm K}$ except for the two closely
spaced contours at $4,640\,{\rm K}$ and $5,220\,{\rm K}$.
For clarity, a cut-plane is used at $33,640\, {\rm K}$ so that the extremely repulsive regions 
for each channel are not plotted. The energy of the CI in Fig. \ref{fig1PES} is $37,700\,{\rm K}$
and the PES for the excited electronic state is not plotted.

The two panels in Figure \ref{fig1PES} depict the interference pathways which can lead
to significant GP effects for each reaction.
In general the total scattering amplitude can be decomposed into contributions from each pathway
labeled by the $1$, $2$, and $3$ in panels (a) and (b).\cite{meadScatter,Althorpescience2005,Althorpe2006,Althorpe2008}
For the inelastic scattering in \ref{fig1PES} (a), pathway \#1 (black) corresponds to 
a non-reactive process and pathways \#2 and \#3 (red) corresponds to an exchange process 
where the two identical D nuclei in each HD channel are exchanged.
For the reactive scattering in \ref{fig1PES} (b), pathway \#1 (black) corresponds to 
a direct reaction process and pathway \#2 (red) corresponds to a looping reaction process.
For the D-atom exchange in \ref{fig1PES} (a), it has been shown that,
due to the direct collinear nature of the reaction,
contributions from pathway \#3 (thin red curve) are negligible
even for high collision energies approaching that of the CI.\cite{kendrickHD2GPJ5,kendrickHD2GPJ,kendrickDH2GPJ,kendrickFeature2003,carlosAlthorpe2005,Althorpescience2005}
Thus, the total scattering amplitude which does not include the 
GP can be written as ${\tilde f}^{\rm NGP} = (1/\sqrt{2})({\tilde f}_1 + {\tilde f}_2)$
where NGP denotes ``No Geometric Phase'' and ${\tilde f}_1$ and ${\tilde f}_2$
are the scattering amplitudes for pathways \#1 and \#2 in Fig. \ref{fig1PES} (a).\cite{meadScatter,Althorpescience2005,Althorpe2006,Althorpe2008}
The GP alters the sign on the scattering amplitude for pathway \#2 
across the branch cut (black dashed curve in Fig. \ref{fig1PES}).
Thus, the total scattering amplitude which includes
the GP is given by ${\tilde f}^{\rm GP} = (1/\sqrt{2})({\tilde f}_1 - {\tilde f}_2)$
where GP denotes ``with Geometric Phase''.\cite{meadScatter,Althorpescience2005,Althorpe2006,Althorpe2008}
The same expressions hold for the NGP and GP scattering amplitudes 
in Fig. \ref{fig1PES} (b).\cite{Althorpescience2005,Althorpe2006,Althorpe2008}
The encirclement of the CI by the combined pathways \#1 and \#2 is obvious in Fig. \ref{fig1PES} (b) but not
so obvious in Fig. \ref{fig1PES} (a).  
Pathway \#2 in Fig. \ref{fig1PES} (a) encircles the CI through ``symmetric encirclement''
(i.e., via the symmetrization of the wave function with respect to permutation of the identical D nuclei).\cite{meadtruhlar79,meadScatter,Althorpescience2005,Althorpe2006,Althorpe2008}

The cross sections and rate coefficients
are computed from the square modulus of the total scattering amplitude
$\vert\vert {\tilde f} \vert\vert = (1/2)\,(f_1^2 +  f_2^2 \pm 2\,f_1\, f_2 \,\cos\Delta )$
where the $+$ and $-$ denote NGP and GP, respectively. 
The complex scattering amplitudes are expressed as ${\tilde f}_1 = f_1\,{\rm exp}(i\,\delta_1)$,
${\tilde f}_2 = f_2\,{\rm exp}(i\,\delta_2)$ and the phase difference $\Delta=\delta_2 - \delta_1$.
If the square modulus of the scattering amplitude for one of the pathways is much larger than the other:
$f_1^2 >> f_2^2$ or  $f_2^2 >> f_1^2$, then the square modulus of the total scattering amplitude
is given by $\vert\vert {\tilde f} \vert\vert\approx f_1^2/2$ or $\vert\vert {\tilde f} \vert\vert\approx f_2^2/2$, respectively,
and the GP effect is negligible.
However, when the squared moduli are similar $f_1^2 \approx f_2^2$, then 
$\vert\vert {\tilde f} \vert\vert \approx f^2\,(1 \pm \cos\Delta)$ where $f=f_1\approx f_2$.
Thus, depending upon the sign and magnitude of $\cos\Delta$, the reactivity can be dramatically enhanced or suppressed.
The maximum interference occurs when $\vert \cos\Delta\vert =1$. 
If $\pm\cos\Delta = +1$ then maximum constructive interference occurs and $\vert\vert {\tilde f} \vert\vert \approx 2\,f^2$,
whereas for $\pm\cos\Delta = -1$, maximum destructive interference occurs and $\vert\vert {\tilde f} \vert\vert \approx 0$.
That is, if $\vert\cos\Delta\vert\approx 1$ then the reactivity can be turned on or off by the sign of the interference term.
Since the GP alters the sign of the interference term, the GP effectively controls the reactivity.

The quantum reactive scattering calculations for the H$_3$ system were done using
a numerically exact time-independent coupled-channel method based on hyperspherical coordinates  
and the GP effect is included using the vector potential approach.\cite{kendrickFeature2003,Pack87,kendrick1996I,Kendrick99,LiYb2015,suppmat}
The computed scattering ($S$) matrices include all open reactant and product 
diatomic vibrational and rotational states on a grid consisting of 112 collision 
energies spanning the range from $1\,\mu{\rm K}$ to $4,640\,{\rm K}$ relative to
asymptotic energy of HD($v=4$, $j=0$) for the H + HD and D + HD reactions, 
and H$_2$($v=4$, $j=0$) for H + H$_2$.
For the highly excited reactant vibrational states, the reaction becomes effectively barrierless 
and exhibits significant reactivity at ultracold collision energies.\cite{cote2011}
The asymptotic energies for HD($v=4$, $j=0$) and H$_2$($v=4$, $j=0$) are $22,109\,{\rm K}$ and
$25,078\,{\rm K}$ relative to the bottom of the asymptotic diatomic potential wells, respectively.
The scattering calculations were carried out using two accurate {\it ab initio} electronic PESs
for the H$_3$ system:  the BKMP2\cite{BKMP2} and the newer one by Mielke, et al.\cite{MielkePES}
\\
\begin{figure}[h]
\includegraphics[scale=0.3]{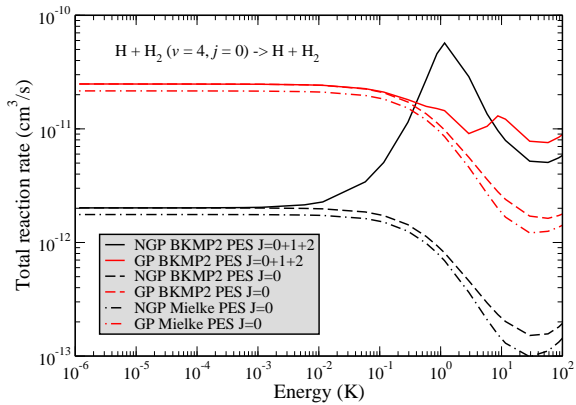}
\caption{
\begin{small} {(color online) The total reaction rate for the H + H$_2$($v=4$, $j=0$) $\to$ H + H$_2$ (para-para) reaction is plotted as 
a function of collision energy (see text for discussion).}
\end{small}
}
\label{fig2}
\end{figure}

\begin{figure}[h]
\includegraphics[scale=0.35]{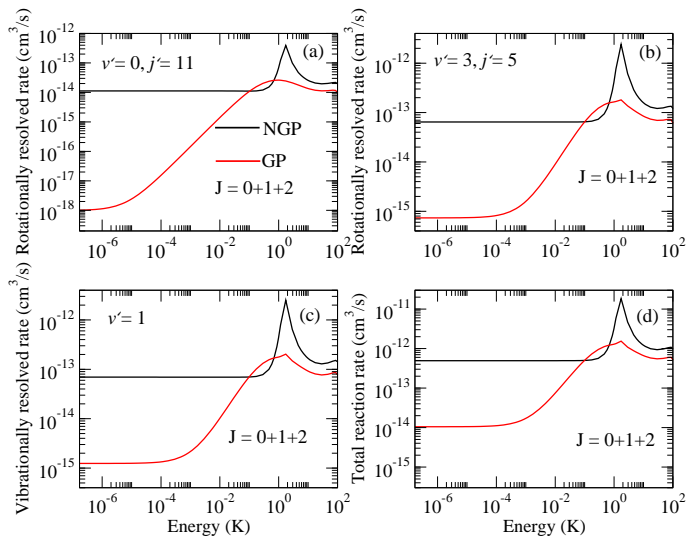}
\caption{
\begin{small} {(color online) Reaction rates for the D + HD($v=4$, $j=0$) $\to$ D + HD($v'$, $j'$) reaction are plotted as a
function of collision energy: (a) $v'=0$, $j'=11$, (b) $v'=3$, $j'=5$, (c) $v'=1$,
and (d) total. The results are for even exchange symmetry and include all values of total angular momentum $J=0-2$ (see text for discussion).} 
\end{small}}
\label{fig3}
\end{figure}

Figure \ref{fig2} plots the total reaction rate for H + H$_2$($v=4$, $j=0$) $\to$ H + H$_2$
summed over all product vibrational and even rotational states (i.e., the para-para transitions). 
The results which include (do not include) the GP are plotted in red (black).
The thick curves include all values of total angular momentum $J=0-2$ and the
thinner curves are for $J=0$ only.  
The rates for each value of $J$ are well converged over the entire energy range.
The total rate is well converged with respect to the sum over $J=0-2$
up to about $2\,{\rm K}$.\cite{suppmat}
The dashed and long-short dashed curves compare the results based on the 
BKMP2 and Mielke PES, respectively.  Both PESs give similar results and predict
that the GP enhances the ultracold reactivity by a full order of magnitude.
Figure \ref{fig3} plots several representative rate coefficients 
for the D + HD($v=4$, $j=0$) $\to$ D  + HD($v'$, $j'$) reaction
using the BKMP2 PES. The rates are computed for even exchange symmetry
(i.e., the nuclear motion wave function is symmetric with respect to 
permutation of the identical D nuclei)
and include all values of total angular momentum $J=0-2$.
The rotationally resolved rate for $v'=0$, $j'=11$ shows that the geometric phase
reduces the reactivity by over four orders of magnitude.
The rate for $v'=3$, $j'=5$ and the vibrationally resolved rate for $v'=1$ 
are reduced by nearly two orders of magnitude when the GP is included.
The total rate which includes the GP is reduced by a factor of 50.
The rates for odd exchange symmetry are similar in magnitude except that the GP 
increases the reactivity in this case.\cite{suppmat}
When both the symmetric and antisymmetric nuclear spin states of D$_2$ are present, 
the scattering results must be summed over both even and odd exchange symmetries including 
the appropriate nuclear spin statistical weights $2/3$ and $1/3$, respectively.
The GP effect is reduced but remains significant (see Table \ref{table1} and Ref. \onlinecite{suppmat}).

Table \ref{table1} lists a representative sample of the ultracold ($1\mu{\rm K}$) reaction rates 
computed for the H$_3$ system both with and without the geometric phase 
and different exchange symmetries.
Most notable are the very large GP effects seen in the rotationally resolved 
rates for the D + HD $\to$ D + HD reaction for each exchange symmetry.
These differences persist, albeit smaller when summed over both exchange symmetries.
The GP effects for the H + HD $\to$ H + HD reaction are overall smaller than
those for the D + HD reaction.
For the H + H$_2$ $\to$ H + H$_2$ para-para reaction, the results have
already been summed over the appropriate nuclear spin states.
Large differences ($\approx$ 10x) between the GP and NGP rates are observed even
when summed over all $v'$ and $j'$.
The total rates summed over all $v'$ and $j'$ using the PES of Mielke et al. are tabulated in the last column.
They are very similar at the state resolved level as well but the overall reactivity 
is slightly reduced (see Fig. \ref{fig2}).
For the D + HD $\to$ H + D$_2$ and H + HD $\to$ D + H$_2$ reactions (not tabulated),
the GP and NGP rates are nearly identical even at the 
rotationally resolved and single exchange symmetry level.
The same applies to the H + H$_2$ para-ortho reaction (not tabulated).
The lack of GP effects for these ultracold reactions is due to the direct collinear
nature of the reaction which results in a tiny contribution from the scattering amplitude
corresponding to looping pathway in Fig. \ref{fig1PES} (b).

\begin{table}
\caption{\begin{small}
Ultracold ($1\,\mu {\rm K}$) reaction rates for the X + HD($v=4$, $j=0$) $\to$ X + HD($v'$, $j'$)
with X=D and H, and the H + H$_2$($v=4$, $j=0$) $\to$ H + H$_2$($v'$, $j'$ even) reactions. The
``evn'' and ``odd'' denote exchange symmetry and the ``GP'' and ``NGP'' denote the results computed
with and without the geometric phase, respectively. The ``(evn) + (odd)'' denote the summed results
over even and odd exchange symmetry.
All rates include the appropriate nuclear spin statistical weights and are in cm$^3$/s.\end{small}}
\begin{tabular}{ccccc}
\hline
Reaction
&{$v'$=0,$j'$=11}
&{$v'$=2,$\sum_{j'}$}
&{Total}
&{Mielke PES}\\
\hline
D + HD(evn) GP & 1.01(-18) & 1.51(-15) & 1.05(-14)& 6.64(-15)\\
\ \ \ \ \ \ \ \ \ \ \ \ \ \ \ \ \ NGP & 1.13(-14) & 1.52(-13) & 4.94(-13)& 2.51(-13)\\
D + HD(odd) GP & 5.48(-15) & 7.52(-14) & 2.44(-13) & 1.25(-13)\\
\ \ \ \ \ \ \ \ \ \ \ \ \ \ \ \ \ NGP & 1.90(-19) & 7.42(-16) & 4.99(-15) & 2.97(-15)\\
\hline
(evn) + (odd)\ GP & 5.48(-15) & 7.67(-14) & 2.54(-13)&1.32(-13)\\
\ \ \ \ \ \ \ \ \ \ \ \ \ \ \ \ \ NGP & 1.13(-14) & 1.53(-13) & 4.98(-13)&2.54(-13)\\
\hline
&{$v'$=0,$j'$=2}
&{$v'$=0,$\sum_{j'}$}
&
&\\
\hline
H + HD(evn) GP & 1.60(-17) & 1.58(-14) & 9.87(-14)&\\
\ \ \ \ \ \ \ \ \ \ \ \ \ \ \ \ \ NGP & 1.22(-14) & 8.80(-14) & 3.15(-13)&\\
H + HD(odd) GP & 3.79(-14) & 2.71(-13) & 9.35(-13)&\\
\ \ \ \ \ \ \ \ \ \ \ \ \ \ \ \ \ NGP & 1.95(-17) & 4.34(-14) & 2.85(-13)&\\
\hline
(evn) + (odd)\ GP & 3.79(-14) & 2.87(-13) & 1.03(-12)\\
\ \ \ \ \ \ \ \ \ \ \ \ \ \ \ \ \ NGP & 1.22(-14) & 1.31(-13) & 6.00(-13)&\\
\hline
&{$v'$=3,$j'$=4}
&{$v'$=3,$\sum_{j'}$}
&
&\\
\hline
H + H$_2$\ \ \ \ \ \ \ \ \ GP & 4.32(-11) & 8.39(-12) & 2.48(-11) & 2.16(-11)\\
\ \ \ \ \ \ \ \ \ \ \ \ \ \ \ \ \ NGP & 2.16(-13) & 5.41(-13) & 2.02(-12) & 1.76(-12)\\
\hline
\end{tabular}
\label{table1}
\end{table}

For ultracold collisions of H or D with a high vibrationally excited 
HD or H$_2$ diatomic molecule,  leading to vibrational quenching,
the reaction pathway is effectively barrierless with
an attractive potential well.\cite{cote2011,miller1980,truhlar1996,jankunas_PNAS2014}
Thus, each scattering pathway in Fig. \ref{fig1PES} 
can be represented by a simple 1D spherical well model.
For this 1D model, the scattering phase shifts are known analytically and 
in the zero energy limit they become effectively
quantized (i.e., they approach $n\,\pi$ where $n$ denotes
the number of bound states in 1D spherical well).\cite{GPHO2cold2015,suppmat}
If the number of bound states in the two different 1D spherical well potentials 
corresponding to the two reaction pathways in Fig. \ref{fig1PES}(a)
differ by an even (odd) number, then $\cos\Delta=1$ ($\cos\Delta=-1$).
Maximum constructive (destructive) interference will occur between
the two scattering amplitudes contributing to ${\tilde f}^{\rm NGP}$,
and the opposite interference behavior will occur for ${\tilde f}^{\rm GP}$.
Thus, the unusually large GP effects reported here 
originate from the isotropic ($s$-wave) scattering and the effective quantization of the scattering phase 
shift which results in $\vert \cos\Delta \vert\approx 1$.\cite{GPHO2cold2015,suppmat}  
The mechanism is general and is expected to hold for many molecules which exhibit CIs
and for which the PES and/or the choice of reactant and product states allows 
for a favorable encirclement.\cite{GPHO2cold2015}

We emphasize that the interference mechanism reported here is a general property
of ultracold collisions and will also occur in molecules without CIs or GP effects.
In general, large interference effects can be expected for barrierless reaction paths
which proceed over a potential well (due to the PES or vibrational excitation) and
include contributions from two interfering pathways (such as reactive and non-reactive).
Experimentally the interference (and hence reactivity) might be controlled by 
the selection of a specific nuclear spin state or by the
application of external electric or magnetic fields to 
(1) alter the relative number of bound states in 
the effective potential wells along each interfering pathway, or (2) alter the
relative magnitude of the two interfering scattering amplitudes. 

\vspace*{-6mm}
\subsection*{Acknowledgments}
\vspace*{-4mm}
\begin{small}
BKK acknowledges that part of this work was done under the auspices of
the US Department of Energy under Project No. 20140309ER of the Laboratory Directed
Research and Development Program at Los Alamos National Laboratory. Los Alamos National 
Laboratory is operated by Los Alamos National Security, LLC, for the National Security
Administration of the US Department of Energy under contract DE-AC52-06NA25396.
The UNLV team acknowledges support from the Army Research Office, MURI
grant No.~W911NF-12-1-0476 and the National Science Foundation, grant No.~PHY-1205838.
\end{small}

\vfill\eject

\clearpage

%
%
\pagebreak

\setcounter{figure}{0}    
\makeatletter
\renewcommand{\fnum@figure}{\figurename~S\thefigure}
\makeatother

\onecolumngrid

\section*{Supplementary Material}
The three-body quantum reactive scattering methodology is based on a time-independent
coupled-channel approach using hyperspherical coordinates\cite{Pack87} and is well suited
for treating ultracold collisions.\cite{LiYb2015}
The method is numerically exact and accurately treats
the body-frame Eckart singularities\cite{Kendrick99} associated with non-zero 
total angular momentum $J$ and includes the geometric phase using
the general vector potential approach.\cite{kendrick1996I,kendrickFeature2003}
A brief summary of the methodology is given here.
In the interaction region where the three atoms are in close proximity (i.e., for small hyperradius $\rho$), 
Smith-Johnson symmetrized hyperspherical coordinates are used.
For larger hyperradius where the reactant and product channels become well defined, 
a properly symmetrized set of Fock-Delves hyperspherical coordinates are used (one for each channel).
The hyperradius $\rho$ is discretized into a set of sectors spanning the range from small to large $\rho$.
The three-body Hamiltonian is diagonalized at each fixed value of $\rho$ to obtain a set of 2D angular wave functions.
The 2D angular solutions are independent of the collision energy so only have to be computed once for
each value of total angular momentum $J$ and inversion parity.
The 2D angular solutions form the basis set for the coupled-channel equations in $\rho$ and are used to compute a 
set of potential coupling matrices within each sector and the overlap matrices between the different 2D solutions at the boundaries of each sector.
The coupled-channel equations are solved for a given collision energy using Johnson's log-derivative propagator method from small to large $\rho$.
Finally, the asymptotic boundary conditions are applied at large $\rho$ to compute the scattering $S$ matrix 
from which the cross sections and reaction rates can be computed.

\begin{figure}[h]
\includegraphics[scale=0.5]{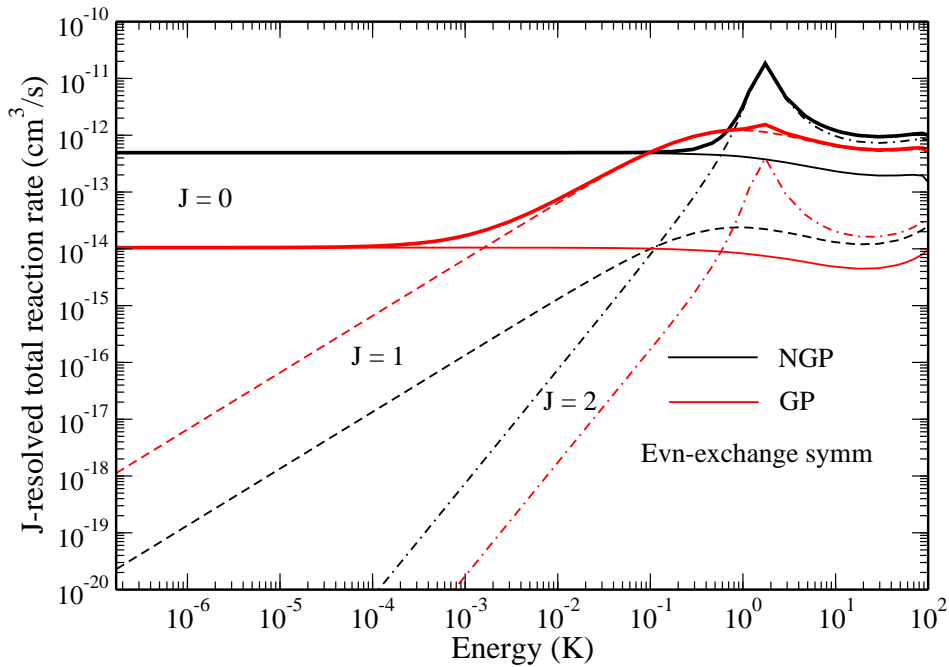}
\caption{
\begin{small} {The individual contributions to the total reaction rate for D + HD($v=4$, $j=0$) $\to$ D + HD
from each value of total angular momentum $J=0$ - $2$ are plotted (see legend).
The results are for even exchange symmetry.  
The GP and NGP reaction rates summed over all values of $J=0$ - $2$ are shown 
as thick red and black curves, respectively (see text for discussion).}
\end{small}
}
\label{figS1}
\end{figure}

The individual contributions from each value of total angular momentum $J=0$ - $2$ 
to the total reaction rate for the D + HD($v=4$, $j=0$) $\to$ D + HD reaction are plotted in 
Fig. S\ref{figS1} as a function of collision energy.  The results with (without) the GP are plotted in red (black).
The thick red and black curves correspond to the total rate summed over all values of $J=0-2$.
The contributions to the rate from $J>0$ rapidly decrease with decreasing collision energy
due to the angular momentum barrier in the entrance channel which increases with $J$.
The rates for each value of $J$ are well converged over the entire energy range.
Fig. S\ref{figS1} shows that the total rate summed over $J=0-2$ is converged for collision energies up to about $2\,{\rm K}$.

\begin{figure}[h]
\includegraphics[scale=0.525]{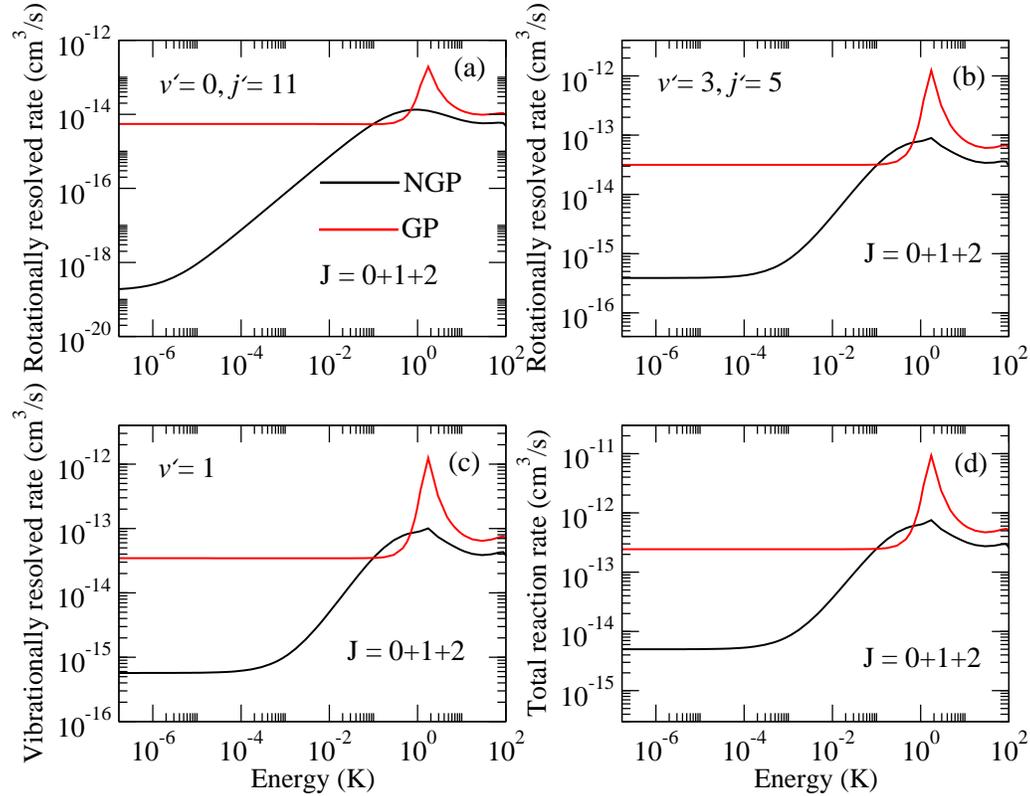}
\caption{
\begin{small} {Reaction rates for the D + HD($v=4$, $j=0$) $\to$ D + HD($v'$, $j'$) reaction are plotted as a
function of collision energy: (a) $v'=0$, $j'=11$, (b) $v'=3$, $j'=5$, (c) $v'=1$,
and (d) total. The results are for odd exchange symmetry and include all values of total angular momentum $J=0-2$.} 
\end{small}
}
\label{figS2}
\end{figure}

\begin{figure}[h]
\includegraphics[scale=0.525]{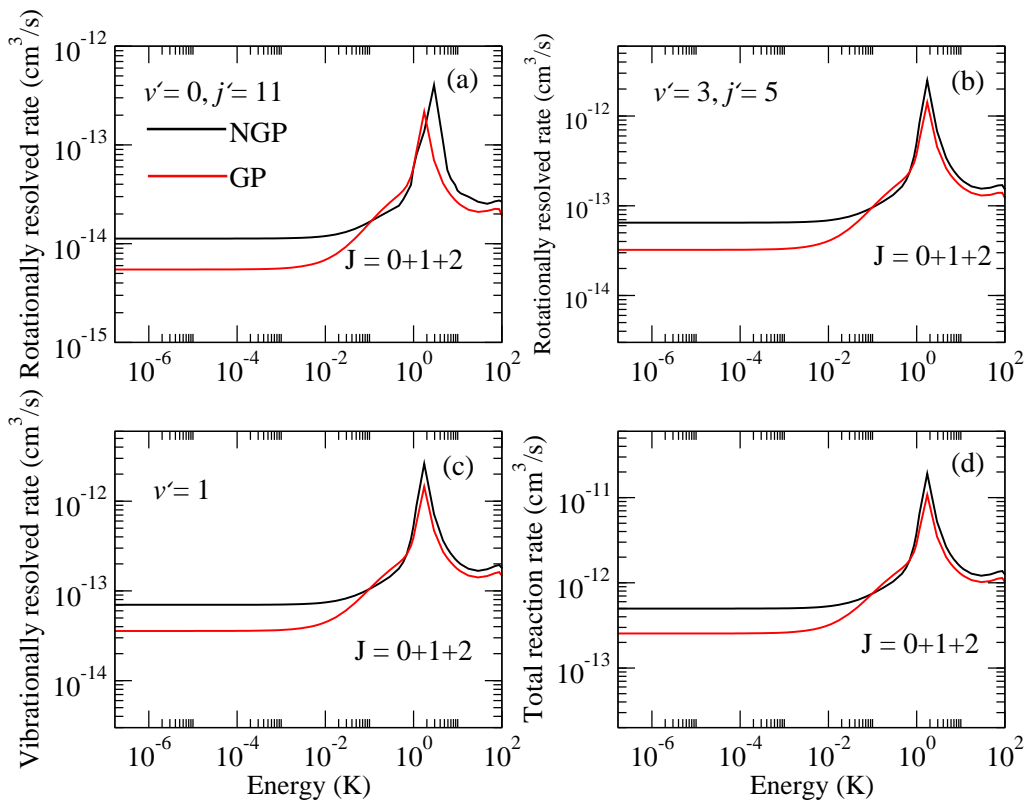}
\caption{
\begin{small} {Reaction rates for the D + HD($v=4$, $j=0$) $\to$ D + HD($v'$, $j'$) reaction are plotted as a
function of collision energy: (a) $v'=0$, $j'=11$, (b) $v'=3$, $j'=5$, (c) $v'=1$,
and (d) total. The results are summed over both even and odd exchange symmetry and include all values of total angular momentum $J=0-2$.} 
\end{small}
}
\label{figS3}
\end{figure}

The reaction rates for the D + HD($v=4$, $j=0$) $\to$ D + HD($v'$, $j'$) reaction are plotted 
in Fig. S\ref{figS2} as a function of collision energy for several values of $v'$ and $j'$ (the same as those in Fig. \ref{fig3} of the main article).
The rates correspond to odd exchange symmetry and include all values of total angular momentum $J=0-2$.
In contrast to the even exchange symmetry results plotted in Fig. \ref{fig3} of the main article, it is the GP ultracold
reaction rates which are largest for odd exchange symmetry.  
Figure S\ref{figS3} plots the reaction rates summed over both even (Fig. \ref{fig3}) and odd (Fig. S\ref{figS2}) exchange symmetries multiplied 
by the appropriate nuclear spin statistical weights.
The differences between the GP and NGP ultracold reaction rates are smaller but still significant.
\\

\begin{figure}[h]
\includegraphics[scale=0.525]{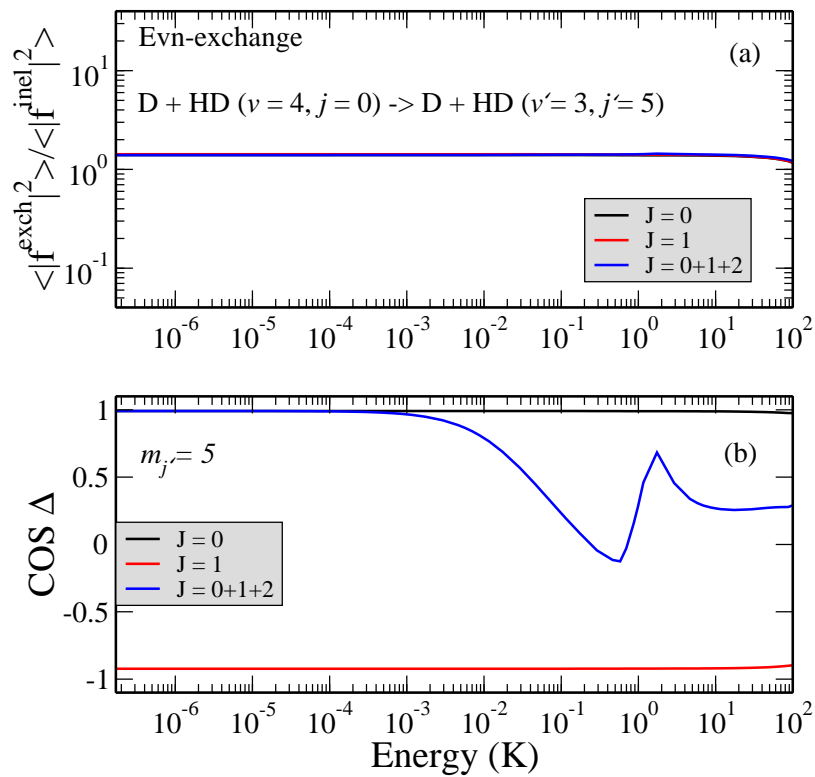}
\caption{
\begin{small} {In panel (a), the ratio of the average square modulus of the exchange (pathway \#2 in Fig. \ref{fig1PES} a) and non-reactive 
(pathway \#1 in Fig. \ref{fig1PES} a) contributions to the total scattering amplitude is plotted as a function of collision energy for the
D + HD($v=4$, $j=0$) $\to$ D + HD($v'=3$, $j'=5$) reaction for even exchange symmetry.  The black and red curves correspond to single values of $J=0$ and $1$, respectively.
The blue curve corresponds to the sum over all values of $J$ between $0$ and $2$. Panel (b) plots the average of $\cos\Delta$ as a function of 
collision energy for $m_{j'}=5$ using the same color designations as in panel (a).}
\end{small}
}
\label{figS4}
\end{figure}

Figure S\ref{figS4} panel (a) plots the ratio of the average square moduli of the exchange and non-reactive pathways 
depicted in Fig. \ref{fig1PES} of the main article
as a function of collision energy for the D + HD($v=4$, $j=0$) $\to$ D + HD($v'=3$, $j'=5$) reaction.
The results for $J=0$, $1$, and summed over $J=0-2$ are plotted in black, red, and blue, respectively.
In each case, the ratio is near unity for a wide range of energies from $1\,\mu{\rm K}$ to $100\,{\rm K}$.
This implies that maximal interference between the exchange and non-reactive pathways depicted in Fig. \ref{fig1PES} (a) is possible,
and that the interference will be governed by the sign and magnitude of $\cos\Delta$.
Figure S\ref{figS4} panel (b) plots the average value of $\cos\Delta$ as a function of collision energy for $m_{j'}=5$
(the results for other values of $m_{j'}$ are similar).
The results for $J=0$ and $1$ show that the scattering phase difference $\Delta$ lies near an even and odd value of $\pi$, respectively.
This leads to maximal interference between the non-reactive and exchange pathways in Fig. \ref{fig1PES} (a).
Since $\cos\Delta=+1$ for $J=0$, the square moduli of the total NGP and GP scattering amplitudes are given by 
$\vert\vert {\tilde f}^{\rm NGP} \vert\vert = f^2\,(1 + \cos\Delta) \approx 2\,f^2$ 
and 
$\vert\vert {\tilde f}^{\rm GP} \vert\vert = f^2\,(1 - \cos\Delta) \approx 0$ which
explains why the NGP ultracold reaction rates are much larger than the GP ones in Fig. \ref{fig3} of the main article.
For $J=1$ the situation is reversed since $\cos\Delta=-1$. This dramatically reduces the differences
between the NGP and GP reaction rates at higher collision energies when the contributions from $J=1$ 
and higher $J$ become important. 
As shown in Fig. S\ref{figS4} panel (b), at the higher collision energies (i.e., $>1\,{\rm mK}$) where
more values of $J$ contribute (blue curve), the $\cos\Delta$ is no longer quantized and tends
to oscillate about zero.
For odd exchange symmetry, the situation is reversed and $\cos\Delta=-1$ and $+1$ for $J=0$ and $1$, respectively.
This explains why the GP ultracold reaction rates are much larger than the NGP ones in Fig. S\ref{figS2}.

\begin{figure}[h]
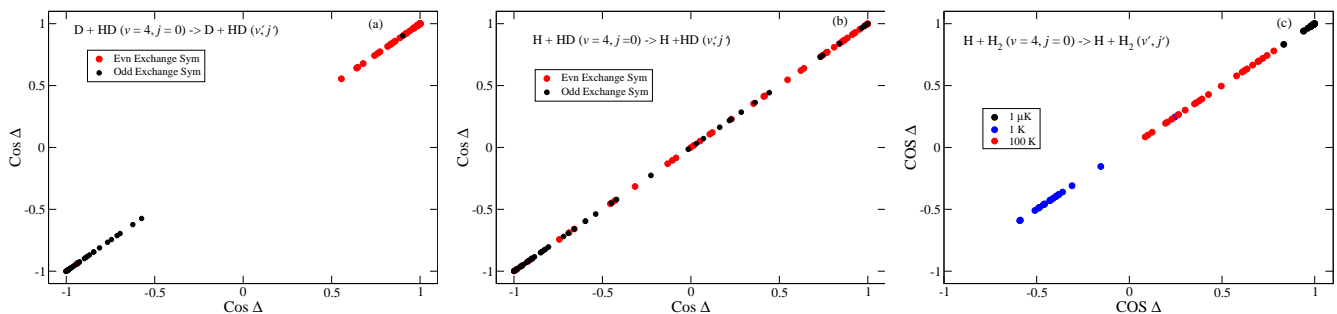

\centering
\begin{tabular}{cc}
\includegraphics[width=0.32\columnwidth]{FigS5_a.eps}\ \ 
\includegraphics[width=0.32\columnwidth]{FigS5_b.eps}\ \ 
\includegraphics[width=0.32\columnwidth]{FigS5_c.eps}\ \ 
\end{tabular}
\caption{
\begin{small} {The $\cos\Delta$ vs $\cos\Delta$ are plotted for three hydrogen exchange reactions. In panel (a), 
the results for D + HD($v=4$, $j=0$) $\to$ D + HD($v'$, $j'$) are plotted at $1\,\mu{\rm K}$.  Each point corresponds
to a $v'$, $j'$ product state and red (black) points correspond to even (odd) exchange symmetry.  Panel (b) is similar
to panel (a) except that the results are for H + HD($v=4$, $j=0$) $\to$ H + HD($v'$, $j'$).
Panel (c) plots the results for H + H$_2$($v=4$, $j=0$) $\to$ H + H$_2$($v'$, $j'$ even) at three different collision energies:
black ($1\,\mu{\rm K}$), blue ($1\,{\rm K}$), and red ($100\,{\rm K}$). }
\end{small}
}
\label{figS5}
\end{figure}

The tendency for $\cos\Delta$ to be near a multiple of $\pi$ (as discussed above in the context of Fig. S\ref{figS4})
at ultracold collision energies occurs for the majority of product states in all three exchange reactions.
Figure S\ref{figS5} plots $\cos\Delta$ vs $\cos\Delta$ at the ultracold collision energy of $1\,\mu{\rm K}$ for the
three exchange reactions: (a) D + HD($v=4$, $j=0$) $\to$ D + HD($v'$, $j'$),
(b) H + HD($v=4$, $j=0$) $\to$ H + HD($v'$, $j'$), and 
(c) H + H$_2$($v=4$, $j=0$) $\to$ H + H$_2$($v'$, $j'$ even).
The preference for $\cos\Delta\approx\pm 1$ is striking.
The value of $\cos\Delta$ for the large majority of product states are clearly clustered at the top-right and bottom-left corners.
In panels (a) and (b), the values $\cos\Delta=+1$ and $-1$ are preferred by the even (red) and odd (black) exchange symmetries, respectively.
The quantization effect is most dramatic for the D + HD reaction (panel a) and the H + H$_2$ reaction (panel c) which explains
why these two reactions exhibit the largest overall GP effects relative to the H + HD reaction (panel b) (see also Table \ref{table1} in the main article).
In panel (c), the $\cos\Delta$ are plotted at three different collision energies: $1\,\mu{\rm K}$ (black), $1\,{\rm K}$ (blue), and $100\,\mu{\rm K}$ (red).
At higher collision energies the $\cos\Delta$ are not quantized and take on a wide range of values with $\vert\cos\Delta\vert < 0.8$.
However, at the ultracold energy of $1\,\mu{\rm K}$ (black), the $\cos\Delta$ become essentially quantized and approach values very close to $+1$.

\begin{figure}[h]
\includegraphics[scale=0.4]{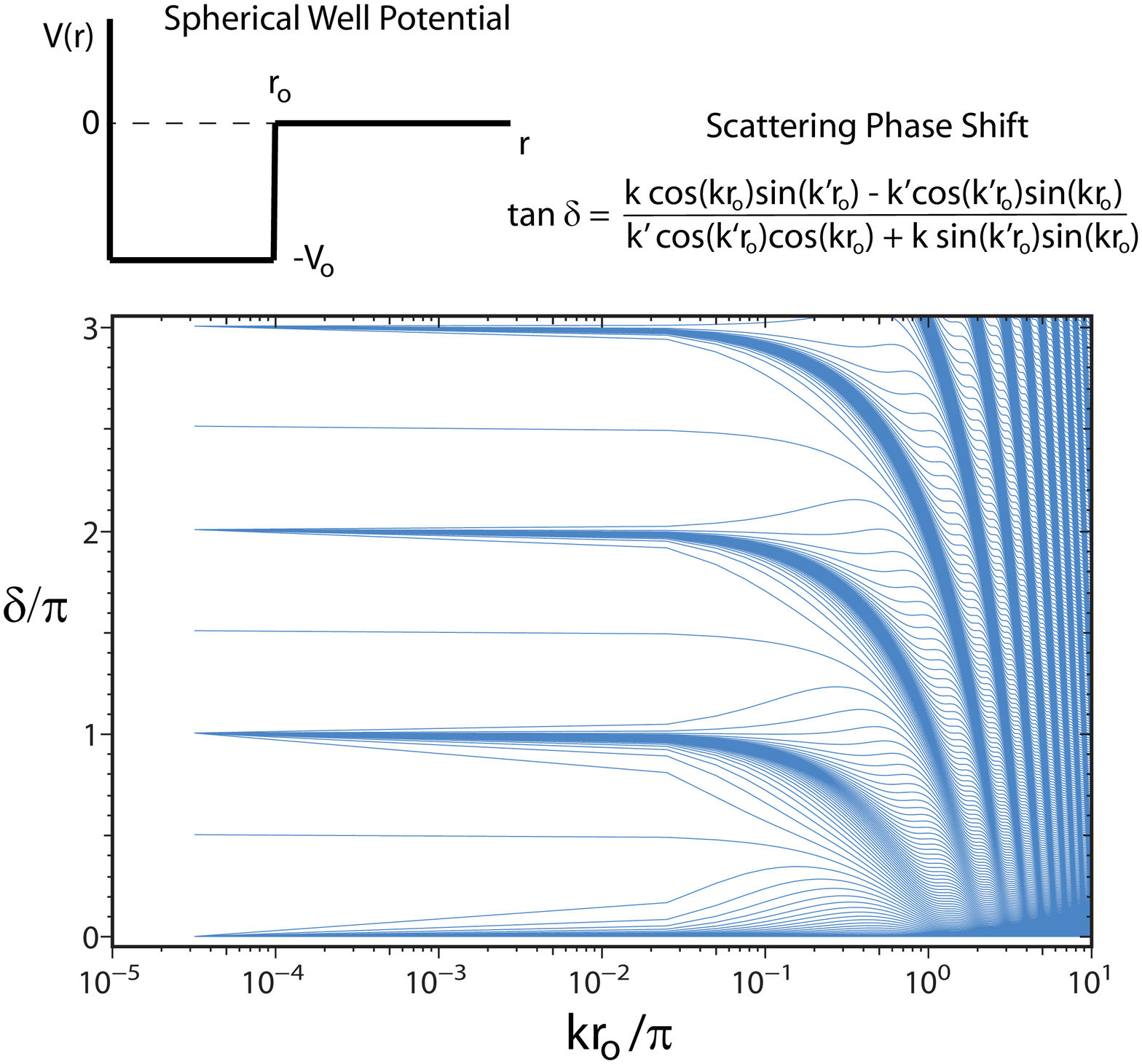}
\caption{
\begin{small} {The 1D spherical well potential model with depth $V_o$ and radius $r_o$, and its associated scattering phase shift (for $l=0$) are plotted.
The scattering phase shifts $\delta$ in units of $\pi$ are plotted as a function of the wave number in units $\pi/r_o$
for different potential well depths $V_o=q^2/(2\,\mu)$ separated by $q=0.025\,\pi/r_o$ (blue curves) where $k'^2 - q^2 = k^2$.  
In the zero temperature limit $k\to 0$, the majority of scattering phase shifts 
approach a multiple of $\pi$ (i.e., they become effectively quantized).}
\end{small}
}
\label{figS6}
\end{figure}

The dynamical origin of the phase quantization observed in Figs. S\ref{figS4} and S\ref{figS5} can
be understood by considering a simple 1D spherical well potential.
Figure S\ref{figS6} shows a 1D spherical well potential with depth $V_o$ and radius $r_o$.
The scattering phase shift $\delta$ can be derived analytically for this model 
and is given by the expression shown in Fig. S\ref{figS6} for $s$-wave scattering (i.e., $l=0$).
The analytic phase shift is plotted as a function of the wave number $k$
and each blue curve corresponds to a different well depth $V_o$.
In the ultracold limit $k\to 0$, the majority of the phase shifts (blue curves) converge towards a multiple of $\pi$.
For very small values of $k$, the phase shift quickly jumps from one value of $\pi$ to another
as the well depth $V_o$ is varied.
The jumps occur at intervals of $\pi/2$ which correspond to resonances associated with 
a continuum state dropping into the well to become a bound state.
In general, the phase shift for $k\to 0$ is equal to $n\,\pi$ where $n$ denotes the number of bound states supported by the potential well.
For high-lying vibrationally excited reactant states, the ultracold hydrogen exchange reaction is effectively
barrierless\cite{cote2011} and proceeds along an effective attractive potential well
(i.e., vibrational adiabat).\cite{miller1980,truhlar1996,jankunas_PNAS2014}
Thus, the scattering phase shifts along each pathway depicted in Fig. \ref{fig1PES} of the main article can be 
modeled using an effective 1D spherical well potential.
Since each pathway has a different effective well depth $V_o$ and radius $r_o$, it will support a different
number of bound states and hence the ultracold scattering phase shift along each pathway will approach a different multiple of $\pi$.
This implies that the phase difference $\Delta$ will also approach a multiple of $\pi$.
If the multiple of $\pi$ is an even or odd number, then $\cos\Delta =+1$ or $\cos\Delta =-1$, respectively.
This ultimately determines whether the interference will be constructive or destructive and whether 
the GP or NGP reaction rates will be the largest.
Furthermore, the effective 1D spherical well potential associated with each of the 
reaction pathways also depends upon the particular reactant and product states involved
(hence the distribution of points observed in Fig. S\ref{figS5}).

\end{document}